\documentclass[10pt]{article}
\usepackage{aas2pp4}
\usepackage{aassize}

\newcommand{\eg}{{\it eg.\,}}
\newcommand{\et}{{\it et~al.\,}}

\begin{document}
\title{First Astronomical Application of a Cryogenic TES Spectrophotometer}
%\vspace{-0.4cm}

\author{R.W. Romani, A.J. Miller, B. Cabrera, E. Figueroa-Feliciano}
\affil{Dept. of Physics, Stanford University, Stanford CA  94305-4060}
\authoraddr{Dept. of Physics, Stanford University, Stanford CA  94305-4060}
%\smallskip
\author{and S.W. Nam}
\affil{NIST, Boulder, CO 80303}
\authoraddr{NIST, Boulder, CO 80303}
\vspace{-0.6cm}

\vspace{-.5cm}

\begin{abstract}
	We report on the first astronomical observations with a photon
counting pixel detector that provides arrival time- ($\delta t = 100$ns) and
energy- ($\delta E_\gamma \le$\,0.15\,eV) resolved measurements from the near IR
through the near UV. Our test observations were performed by coupling this
Transition Edge Sensor (TES) device to a 0.6m
telescope; we have obtained the first simultaneous optical
near-IR phase-resolved spectra of the Crab pulsar. A varying infrared 
turnover gives evidence of self-absorption in the pulsar plasma.
The potential of such detectors in imaging arrays from a space platform
are briefly described.
\end{abstract}

\section{Introduction and Experimental Apparatus}

	There is growing interest in using cryogenic bolometer
technology in array detectors that can provide imaging spectrophotometry
with high quantum efficiency in the optical range. In particular, there are well
established programs developing Superconducting Tunnel Junction (STJ)
devices (\eg Peacock, \et 1996) for such arrays. Manufacture and control of STJs
have proved particularly challenging, but the wide range of potential applications
(\eg Jakobsen 1999), has motivated substantial investment in this technology.
We have recently demonstrated a second bolometric device for time- and energy-
resolved photon detection spanning the near IR through UV (Cabrera, \et 1998).
This Transition-Edge Sensor(TES) device, based on the sharp
superconducting-normal transition in tungsten (W) thin-films offers remarkable
new capabilities for problems requiring low energy resolution 
time-resolved spectrophotometry at low light levels. We have tested this
instrument by coupling to small telescopes at our suburban campus site. Initial
test observations, including first measurements (12/98) of the Crab pulsar were
reported in Romani, \et (1998); here we describe in more detail our test system
and new optical/IR phase-resolved pulsar spectra. Very recently, the ESA STJ group
has also detected the Crab pulsar in the optical using a focal plane STJ
array at the 4.2m WHT (Perryman, \et 1999), albeit with lower statistics and
energy resolution than achieved here.

	In the present application, our detectors are 18 micron squares
of 40\,nm thick tungsten (W) patterned on silicon. The Si substrate is 
cooled below the
$T_{C} \approx 80\,$mK transition in a Kelvin Ox 15 dilution refrigerator.
By resistively heating (Joule power $\propto V^2/R$) the pixels at a fixed voltage,
the W is stably driven into the transition range; a photon
is detected as a temporary deficit in the required Joule power to
keep the pixel at the stable set point.  This 
\newpage
\mbox{}
\begin{minipage}[b]{3in}
\vspace{3.75in}
\end{minipage}
\hspace{-0.81cm}
`electro-thermal' feedback
speeds up recovery to the quiescent power level (Irwin 1995). Pixels with time 
constants 
as short as 2\,$\mu$s have been produced, allowing photons to be counted at
rates of $\sim 30$\,kHz per channel with minimal pile-up. The current
deficit pulse associated with arrival of a photon is amplified via
a DC SQUID array (Welty \& Martinis 1993), digitized, assigned a peak 
height, time-stamped using a GPS receiver and recorded to computer disk.

	Tungsten is grey and photons hitting the pixel are absorbed
with an efficiency varying from $\ga 50$\% in the optical/UV to $\sim
10$\% in the near IR (Palik 1985). This substantial intrinsic efficiency can, 
in principle,
be increased to near unity by `blackening' the surface with a passive absorber,
{\it e.g.}, Au-black (Harris 1956). Calibration photons were introduced into the 
refrigerator dewar via optical fiber -- we have demonstrated single photon
detection from 0.3\,eV (4\,$\mu$m) to the fiber cut-off at $\sim$3.5\,eV 
(0.35\,$\mu$m);
the intrinsic device sensitivity continues to the saturation energy of
10-15\,eV. Such an ultra-broad band is accessible only from space where these
superconducting devices offer unprecedented new sensitivity for a number
of important astronomical observations.
We have initiated
a series of ground-based experiments to illustrate the potential of these
instruments.

\begin{figure}[!h]
%\plotfiddle{figures/tschem.ps}{8.5truecm}{90}{20}{20}{500}{500}
\plotone{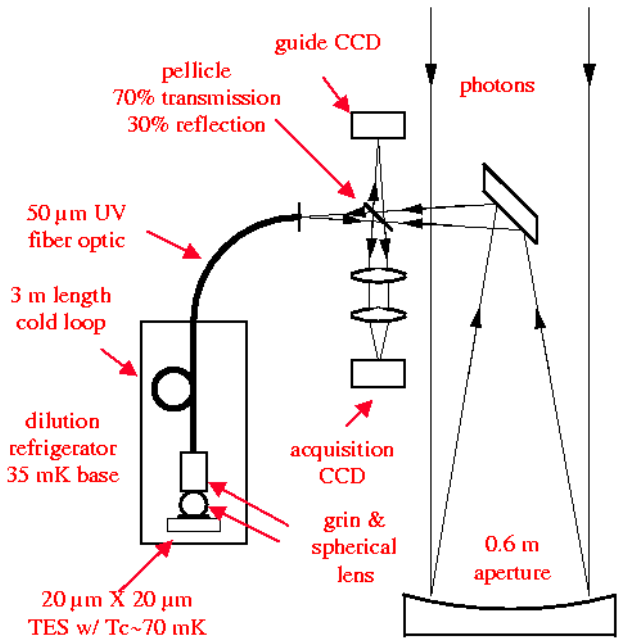}
\caption{Schematic of our TES fiber coupled system at the Stanford Observatory 0.6m.}
\label{schem}
\end{figure}

	The substantial IR sensitivity of the detector, coupled with 
rapid increase in the atmosphere and telescope thermal emissivity in the 
near IR means that careful filtering to exclude $E_\gamma < 0.7$\,eV 
($\lambda > 1.7~\mu$m) photons was required to avoid detector
saturation. Since conventional filter glasses typically absorb beyond
$\sim 2.5~ \mu$m and bandpass filters transmit in discrete octaves, we
have adopted a novel filtering scheme to demonstrate the sensitivity
to a larger energy range than that of CCDs or conventional
IR detectors. As radiation is coupled via optical fiber, we have spooled
3\,m of `wet' (high OH) fiber at the 1\,K stage. The OH in
this fiber provides strong attenuation (Humbach 1996) below 0.7\,eV and in secondary 
bands (e.g. 0.91\,eV) where, in any case, atmospheric OH limits transmission 
of astrophysical signals.  The system
was thus sensitive from the near-IR J and H bands through the visible.
Using a 50\,$\mu$m core fiber we have coupled TES sensors to a 0.6\,m 
telescope at our University teaching observatory (Fig 1). The fiber, whose core
subtended 4.8$^{\prime\prime}$ at the f/3.5 Newtonian focus, was mounted
in a reflective decker and was viewed through a pellicle by an acquisition 
camera.  At the cold stage, the fiber output 
was focused through a ball and GRIN lens system to improve matching to
isolated 18\,$\mu$m square pixels. As focusing was incomplete, $\sim$25\% of the
photons fell on the (unmasked) voltage bias leads of the pixels 
(Cabrera, \et 1998),
for which incomplete energy transfer to the W produced pulses with $\le 0.5$ 
of the main photon energy. The 
contribution of these `rail hits' to the energy PSF appears
in the monochromatic calibration photon spectrum in Figure 4.
Photon pulses were digitized with 6 bit 
resolution, time-stamped to 0.1\,$\mu$s precision and recorded for later 
analysis.

	Observations of the faint visible/IR spectrophotometric standards 
HD 73616 and HD 284504 (Gunn and Stryker 1983) allowed an estimate of 
the system efficiency. The atmosphere and fiber plus detector system 
showed $\sim 2\%$ peak effective QE with a response spectrum consistent with 
the known atmosphere, fiber and W absorptivity. The bulk of the
loss is traceable to known fiber inefficiencies in coupling
into the dewar; with improvements to this system, we expect to realize the high 
intrinsic sensitivity of the W absorber.

\section{Crab Pulsar Spectrophotometry}

	On the nights of January 9-11 1999, we observed 
several sources including the Crab pulsar PSR B0531+21 to illustrate the
time and energy capabilities of the device. Despite poor conditions
(high cirrus, $\sim 3^{\prime\prime}$ imaging, moon, and suburban light 
pollution) the data obtained clearly illustrate the potential of 
this new technology.
\begin{figure}[!h]
\plotone{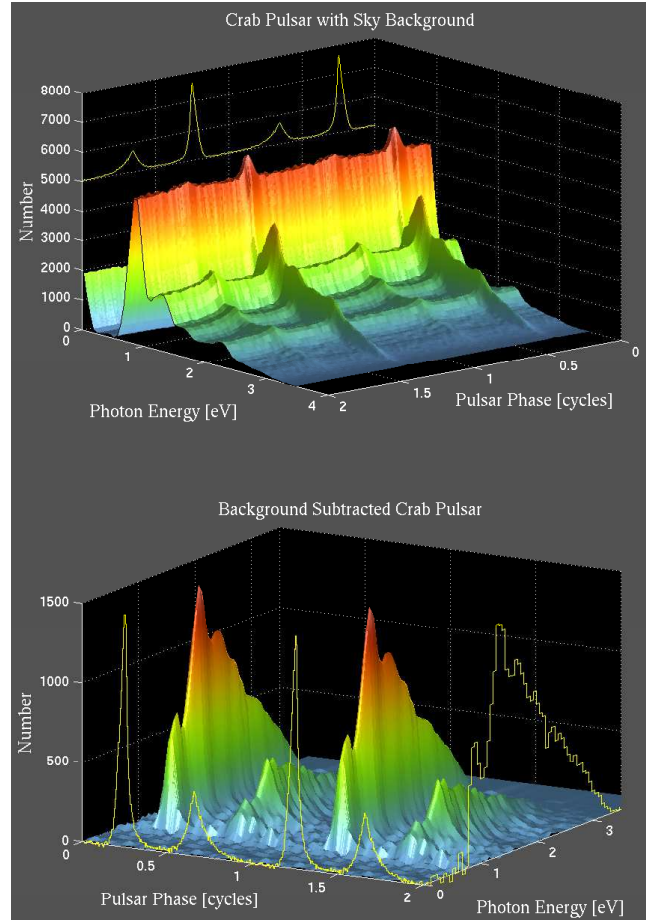} %{8.5truecm}{0}{20}{50}{500}{500}
\caption{Phase resolved spectra of the $P\approx 33.5$\,ms Crab pulsar 
optical/IR emission. Above, the rail-hit corrected count spectra 
folded into 200 phase bins. Below, after off-pulse sky subtraction.}
\label{2Dbacksub}
\end{figure}

	The photons were barycentered using the current Jodrell Bank radio 
ephemeris (Lyne, Pritchard and Roberts 1999) and monitored for signal 
(guiding and atmospheric transmission)
and background (cirrus, moon and local lighting) fluctuations. After 
moderate S/N cuts, some $10^6$ Crab pulsar photons were accumulated 
into spectra in pulsar phase bins of width 0.005 (168\,$\mu$s, Figure 2).
Figure 3 shows our folded light curves in visible 
and near-IR bands. This observation is, to our knowledge, the first simultaneous
detection of the Crab at these wavelengths. The curves, normalized at the
pulse wings below the half power-point, show a substantial (20\%) suppression 
of the main pulse peak, as noted earlier in lower time-resolution data (Penny 1982).
From our light curves, we find that the shape of the decrement is well fitted 
by synchrotron self-absorption with an optical depth
$\tau \approx 2.7 (5.1\times 10^4)^{-\alpha} (f_E/30{\rm \,mJy}) (D_2/l_7)^2 
(\phi\, B/B_{LC})^{0.5} E_{eV}^{-2.5}$, where the pulsar peak has a
power-law flux spectrum $f_E \sim 30 E_{ev}^\alpha\,$mJy with $\alpha$ breaking
from $\approx -0.3$ in the UV to $\approx 0$ at 2\,eV. This flux is emitted
from a region of size $10^7l_7\,$cm near the light cylinder where the
field is $B_{LC} \approx 8 \times 10^5\,$G. We assume a
distance of $2D_2\,$kpc. With these estimates the optical depth fit from
our IR band light curve with $\alpha=0$ gives an $e^\pm$ pitch angle
$\phi \sim 2 \times 10^{-3}$. Comparison with flux measurements at longer 
wavelengths (Middleditch, Pennypacker and Burns 1983) indicates that the 
phase-averaged spectral break is likely
caused by a low energy cut-off in the $e^\pm$ spectrum producing a critical
frequency  $E_c\approx 2\,$eV. With our estimated $B$ and $\phi$, the pair spectrum
then cuts off at $\Gamma_{e^\pm} \approx 400$. Gap models (Romani 1996)
predict that different pulse phases arise from regions with varying $B$ and
plasma parameters. For example, the second pulse emission stems from lower altitude 
where a larger dipole $B$ is compensated by a smaller initial 
$\phi$ and rapid cooling of the $e^\pm$ to lower $\Gamma$. Higher S/N 
data from $0.5-3\,$eV following $\alpha$, $E_c$, and $\tau$ through the
pulse would allow a map of field and plasma variations through the magnetosphere.
We have not detected here IR spectral asymmetries noted
(Eikenberry 1997) in the rising and falling
slopes of the pulses, but this may be ascribed to our modest IR count statistics.
Again higher S/N observations showing these effects would be of interest, as their
interpretation is sensitive to the precise IR/optical phasing.
\begin{figure}[!b]
\plotone{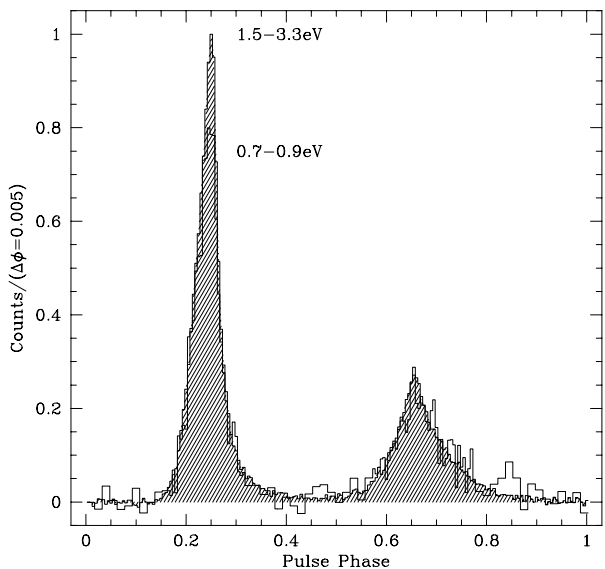}
\caption{Optical and near-IR pulse profiles, showing the
$\sim 20$\% self absorption of the main pulse below 1\,eV.}
\label{lc}
\end{figure}

\begin{figure}[!h]
\plotone{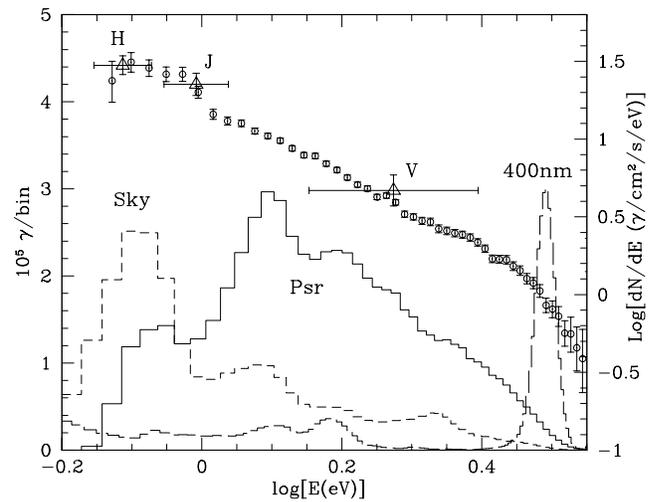}
\caption{Count spectra from a calibration monochromater line (400\,nm), along
with rail-hit subtracted spectra from the main pulse (Psr, $\phi=0.215-0.265$)
and off-pulse (Sky, $\phi=0.875-0.125$) intervals. The flux calibrated 
main pulse (circles) is compared with previously measured  (Middleditch, 
Pennypacker and Burns 1983)
main pulse colors (right scale).}
\label{spectrum}
\end{figure}

	Our flux calibrated spectrum (Figure 4) shows good agreement with 
published (Middleditch, Pennypacker \& Burns 1983) 
near IR and visible main pulse
photometry, although systematic errors in measurement 
of the calibration stars contribute residuals to the expected 
synchrotron spectrum exceeding those of count statistics.
In addition to measuring spectral variations 
through the full period (Figure 2), the timing of individual photons
allows the search for other periodicities and stochastic fluctuations.
For example, 60\,Hz line fluctuations are seen in the $\sim 2.1\,$eV Na
background 
emission line dominated by street lamps in the San Francisco bay area.
We can also examine photon arrival time statistics: our data show
that both the counts per period in the main pulse and the distribution of
photon arrival times within each pulse are consistent with Poisson at
the 95\% confidence level, for
intervals larger than the maximum $12\mu$s acquisition system dead time.

\section{Future Potential}

	Several improvements to the existing TES sensor
have been demonstrated or designed. Count rates up to 30\,kHz per read
channel have been measured with appropriate tuning of the device 
transition temperature. Rail masks to eliminate the low energy tail of the
PSF can be used to ease IR spectral analysis. Thermal isolation of the W on
a Si$_3$N$_4$ membrane should additionally improve energy resolution 
to $\sim 0.05\,$eV. Multiplexing schemes have been tested (Chervenal, \et 1999) which
will allow arrays substantially larger than the few pixel/few read-channel
capability of the present system; $\sim$100$\times$100 arrays may be achieved
with present multiplexing designs, while advanced schemes are being explored
which may allow readout of a programmable subset of megapixel-scale arrays.

\begin{figure}[!b]
\plotone{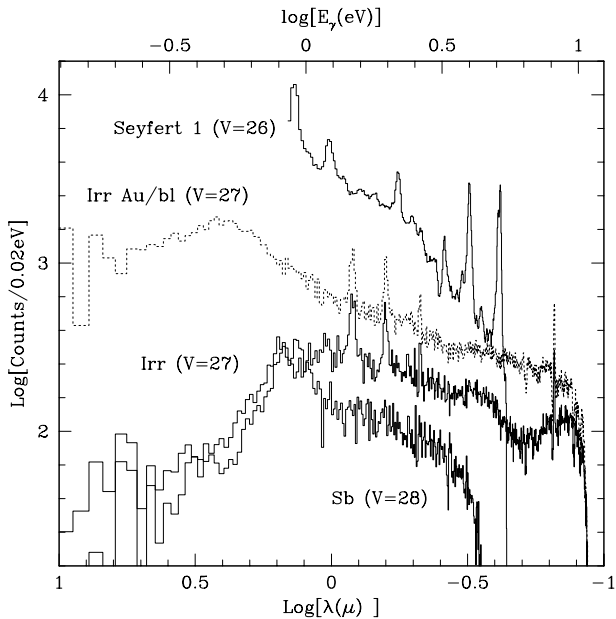}
\caption{Simulated background subtracted spectra from one 
0.1$^{\prime\prime}$ pixel of a TES array on a 1\,m aperture space telescope. 
Passively cooled optics, zodiacal background from moderate ecliptic latitude,
0.05\,eV resolution and 10\,h exposure are assumed. Sample synthetic spectra 
(Fioc \&  Rocca-Volmerange 1997) show
a portion of an Sb galaxy ($M=10^9 M_\odot$, $z=2$)
a low mass irregular galaxy ($M=1.5 \times 10^9 M_\odot$, age=30My, $z=0.25$)
and a low luminosity Seyfert 1 nucleus (scaled from
NGC5548 (STScI AGN atlas); $m_V=26$, $z=1$). The dashed line shows the increased S/N
expected from coating the bare W with an absorbing ({\it e.g.}, Au-black) film.
Continuum breaks and a number of lines are well detected; broad ({\it e.g.}, AGN)
lines are resolved in the blue.}
\label{simulation}
\end{figure}

	Our test data already illustrate the potential of a TES 
spectrophotometer for studies of pulsars and stochastically fluctuating 
({\it e.g.}, interacting black 
hole and neutron star binary) compact object sources. Important
experiments probing emission zone structures, pulsar plasma physics and 
relativistic photon propagation follow immediately from observations with
the existing system at large aperture telescopes. Even modest scale arrays
should enable further applications in many fields.
For example in biochemical studies time- and energy-resolved fluorescence 
imaging of multiply labeled biomolecules can characterize configuration
and chemical activity on $\mu$s timescales. 
% (S. Chu, priv. comm.). 
A number of important applications for astrophysics and cosmology have also
been noted (Jakobsen 1999) for such detectors. The most striking progress
will, of course, stem from placement of a TES array 
system in space, where diffraction-limited pixels and cool optics 
allow {\it simultaneous} observation over some two decades of photon 
energy. To illustrate, Figure 5 shows simulated count spectra with a small 
(1\,m) aperture and modest exposure. The modeled sources are comparable
to the {\it faintest} galaxies and galaxy components detected in the
Hubble Deep Field (Williams, \et 1996). TES areal spectrophotometry should
enable a whole new class of studies, providing
morphology, redshift, spectral classification, and chemical evolution studies
at the present limits of visibility in the universe.

\begin{acknowledgements}
This work was supported in part by grants from NASA (NAG5-3775 and NAG 5-3263),
from the DOE (DE-FG03-90ER40569) and from
the Research Corporation; devices were fabricated in the Stanford
Nanofabrication facility. We thank Kent Irwin and
John Martinis of NIST for continued collaboration on TES and SQUID technology.
\end{acknowledgements}

\smallskip

\end{document}